\documentclass[12pt,reqno]{amsart}
\usepackage{color}
\usepackage[toc]{appendix}

\usepackage{amsfonts,amsmath}
\usepackage{latexsym}
\usepackage{stmaryrd}
\usepackage{amssymb}
\usepackage{amsmath,amssymb,amsthm,hyperref,mathrsfs}
\usepackage{graphicx}
\usepackage{times}

\usepackage{latexsym}

\usepackage{t1enc} 
\newtheorem{Theorem}{Theorem}[section]
\newtheorem{Proposition}[Theorem]{Proposition}
\newtheorem {Cor}[Theorem]{Corollary}
\newtheorem {pro}[Theorem]{Proposition}

\newtheorem {Lemma}[Theorem]{Lemma}
\newtheorem {rem}[Theorem]{Remark}
\newtheorem {rems}[Theorem]{Remarks}
\newtheorem {com}[Theorem]{Comment}
\newtheorem {coms}[Theorem]{Comments}
\newtheorem {notation}[Theorem]{Notation}
\newtheorem {Definition}[Theorem]{Definition}
\newtheorem {exam}[Theorem]{Example}

\newcommand{\bcom}{\begin{com} \rm } \newcommand{\ecom}{\end{com}}
\newcommand{\bcoms}{\begin{coms} \rm } \newcommand{\ecoms}{\end{coms}}
\newcommand {\bdef}{\begin{Definition}}
\newcommand {\edefi}{\end{Definition}}
\newcommand {\bl}{\begin{Lemma}}
\newcommand {\el}{\end{Lemma}}
\newcommand {\bethe}{\begin{Theorem}}
\newcommand {\eethe}{\end{Theorem}}
\newcommand {\bp}{\begin{pro}}
\newcommand {\ep}{\end{pro}}
\newcommand {\bcor}{\begin{Cor}}
\newcommand {\ecor}{\end{Cor}}
 \newcommand {\brem }{\begin{rem} \rm }
\newcommand {\erem }{\end{rem}}
 \newcommand {\brems }{\begin{rems} \rm }
\newcommand {\erems }{\end{rems}}

\newcommand{\ba}{\begin{array}}
\newcommand{\ea}{\end{array}}

\newcommand {\be}{\begin{equation}}
\newcommand {\ee}{\end{equation}}
\newcommand {\bde}{\begin{displaymath}}
\newcommand {\ede}{\end{displaymath}}
\newcommand {\beq}{\begin{eqnarray*}}
\newcommand {\eeq}{\end{eqnarray*}}
\newcommand {\beqa}{\begin{eqnarray}}
\newcommand {\eeqa}{\end{eqnarray}}

\def\proof{\noindent {\it Proof. $\, $}}
\def\finproof{\hfill $\Box$ \vskip 5 pt}
\def \ind{1\!\!1}

\newcommand{\bG}{\mathbb{G}}
\newcommand{\bF}{\mathbb{F}}

\newcommand{\bP}{\mathbb{P}}
\newcommand{\bQ}{\mathbb{Q}}
\newcommand{\bR}{\mathbb{R}}

\newcommand{\E}{\mathbb{E}}

\newcommand{\F}{\mathcal{F}}
\newcommand{\G}{\mathcal{G}}

\makeatletter
\newcommand*\bigcdot{\mathpalette\bigcdot@{.5}}
\newcommand*\bigcdot@[2]{\mathbin{\vcenter{\hbox{\scalebox{#2}{$\m@th#1\bullet$}}}}}
\makeatother

\begin{document}

\title[Some No-Arbitrage Rules for Converging Asset Prices]{Some No-Arbitrage Rules for Converging Asset Prices under Short-Sales Constraints}

\author{Delia Coculescu}
\address{University of Zurich\\ Departement of Banking and Finance, Plattenstrasse 32\\ Z\"{u}rich 8032, Switzerland.}
\email{delia.coculescu@bf.uzh.ch}
\author{Monique Jeanblanc}
\address{Laboratoire de Math\'ematiques et Mod\'elisation d'\'Evry (LaMME),   UMR CNRS 8071,  Univ  Evry, Universit\'e Paris Saclay,
.} \email{monique.jeanblanc@univ-evry.fr}
\begin{abstract} Under short sales prohibitions, no free lunch with vanishing risk (NFLVR-S) is known to be equivalent to the existence of an equivalent  supermartingale measure for the price processes (Pulido \cite{Pulido14}). For two given price processes, we translate the property (NFLVR-S) in terms of so called structure conditions and we introduce the concept of fundamental supermartingale measure. When a certain condition necessary to the construction of the fundamental martingale measure is not fulfilled, we provide the corresponding arbitrage portfolios.  The motivation of our study lies in understanding the particular case of converging prices, i.e., that are going  to cross at a  bounded random time.
\end{abstract}
\maketitle

\section{Introduction}
In arbitrage-free financial markets, the law of one price simply states that \textit{similar} financial assets, i.e., that have identical payoffs, should be sold at the same price in different locations. There are some particular assumptions about the financial markets that lead to this fundamental result, importantly investors need to be able to observe the prices in the different locations and to sell short the corresponding assets. Also, there should be no transaction costs. Indeed, under these assumptions, any investor is able to construct an arbitrage portfolio consisting in a short position in the (relatively) overpriced asset and a long position in the (relatively) underpriced asset, thus making an immediate profit. This represents the simplest arbitrage strategy one can encounter: not only is it a buy and hold strategy, but additionally, it is model independent, i.e., does not rely upon an underlying model for describing the prices dynamics in time.

Obviously, in case of short sales prohibitions, the above described arbitrage portfolios are impossible to construct, hence similar assets may have differing prices: the rule of one price does not apply. A question arises naturally: How may the differing    prices behave as stochastic processes within the limits of no arbitrage with short sales constraints? The aim of this paper is precisely to shed light on this question. For this, we are going to introduce the notion of  converging prices, that is,  price processes that are expected to "cross", i.e., to reach almost surely  the same value over some bounded horizon, which is the mathematical description of the similar assets.

 We study the probabilistic properties of such processes when one imposes the no free lunch with vanishing risk condition under short sales constraints, abbreviated  (NFLVR-S). This condition  was introduced by Pulido in \cite{Pulido14}, as the counterpart -when investors are not allowed to short sell- of the no arbitrage paradigm (NFLVR)  of Delbaen and Schachermayer (see \cite{DelbScha94} and \cite{DelbScha98}). For the reader's convenience, the definition of (NFLVR-S) is provided in Section \ref{sec::TheModel}.

Based on  previous work by Jouini and Kallal \cite{JouiKall95}, Fritelli \cite{Fritelli97}, Pham and Touzi \cite{PhamTouz99}, Napp \cite{Napp03} and Karatzas and Kardaras \cite{KaraKard07}, the paper by  Pulido  \cite{Pulido14} establishes important properties of price processes under short sale prohibitions namely the equivalence between (NFLVR-S) and the existence of an equivalent supermartingale measure for the price processes.
In the current paper, we shall rather translate the condition (NFLVR-S) in terms of so-called "structure conditions" for two underlying stochastic processes.

In the framework of converging prices, the existence of imperfect and asymmetric information is crucial to justify the formation and persistence in time of the differing prices. This element is integrated in our analysis: we assume that each individual price is formed given some distinct information set (filtration) a priori unrelated with the information set that drives the price formation in a different location, except measurability of the final payoff in both situations. We then analyse the no arbitrage conditions from the perspective of an agent (called the insider) that has access to a global information set, i.e., that comprises the observation of the two differing prices. The insider can trade in both markets, but cannot sell short.

 There are many examples of converging prices, the simplest being a future contract and its underlying asset, or the two portfolios arising from the call-put parity (i.e., one consisting of a call option and bonds, the second of a put option and underlying stock).  In  markets with  short sales prohibitions, the call-put  parity  is not expected to hold in every point in time but we observe the identity of the payoffs at maturity.  Other examples of convergence are represented by  some portfolios that are commonly used in capital structure arbitrages or the pairs trading. Note however that in these cases the convergence is model-based; in capital structure arbitrages a particular "structural" model is assumed to explain the joint evolution of the prices for the different securities with common issuer, while in the pairs trading, the pairs are selected upon a statistical analysis. Nevertheless, assuming that the underlying models are "correct" the questions remains the same: how to construct the strategies when selling short is not possible?
Finally, our framework applies well to the case of similar derivative contracts that are sold over the counter, and thus differing prices typically arise as a consequence of a imperfect information between the different sellers and buyers.

The remaining of the paper is organised as follows: Section
\ref{sec::TheModel} introduces the probabilistic model for the two
converging prices and recalls the no arbitrage framework we adopt in this paper. Section \ref{Section3} establishes the ''structure conditions'' in
Theorem \ref{ThmGenRepr}.   In Section \ref{sec::MainResult}, we
derive sufficient conditions for the existence of a
supermartingale measure as well as some necessary conditions. We introduce a probability measure that we call fundamental supermartingale measure and arbitrage portfolios are provided when the fundamental supermartingale
measure does not exist.  Section \ref{sec::Examples}
analyses many examples of converging prices. Let us emphasise that
our main results, i.e., Theorem  \ref{ThmGenRepr} and Theorem
\ref{MainThm}, are more general: the property of the two prices
to be converging is not used for deriving these results.

\section{A Stochastic Model with two Converging  Asset Prices }\label{sec::TheModel}

In this paper, all probabilities and filtrations are defined on a probability space $(\Omega, \F,\bP)$.

We consider two financial assets, possibly traded in different locations (exchanges). Their respective price processes denoted by $X:=(X_t)_{t\geq 0}$ and $Y:=(Y_t)_{t\geq 0}$ are modelled as positive stochastic processes  on  $(\Omega, \F,\bP)$.
We denote by  $\bF^X:=(\F^X_t)_{t\geq 0}$ (resp. $\bF^Y:=(\F^Y_t)_{t \geq 0}$)  the right-continuous  $\bP$-augmented filtrations
 of $X$ (resp. $Y$).    For simplicity, we suppose that the spot interest rates are constant and equal to zero, that is, the price processes $X$ and $Y$ are already discounted.

 We shall assume that the dynamics of the two prices reflect a (local) equilibrium, namely there exist equivalent martingale measures for each asset individually, when considered as stochastic processes in their own filtration:
\begin{itemize}
\item[(NA-X)] There exists $\bQ^X\sim \bP$ such that the price process $X$ is an $(\bF^X,\bQ^X)$-local martingale (in other words, $\bQ^X$ is a martingale measure for $X$ in its own filtration).
\item[(NA-Y)] There exists $\bQ^Y\sim \bP$ such that the price process $Y$ is an $(\bF^Y,\bQ^Y)$-local martingale.
\end{itemize}
Note that this implies that $X$ and $Y$ are $\bP$ semimartingales in their own filtrations.
 We shall work with the right-continuous versions for $X$ and $Y$.

Notice that the assumptions above exclude price processes that are
predictable and of finite variation in their own filtration. This
pattern would represent a trivial case to examine, so we do not
lose much by excluding it. However, the pattern (i.e., price
processes that are predictable and of finite variation) can still
appear as we shall consider the price processes in a larger
filtration.

We assume that an investor (called hereafter the insider) is able
to observe the price dynamics in the two locations, so that
his  information flow is given by  $\bG:=(\G_t)_{t\geq 0}$
with
\[
\G_t= \cap _{s>t}\F^X_s\vee \F^Y_s.
\]
Also, the investor has a bounded trading horizon, denoted $T$, which is a $\bG$-stopping time.

Many examples that we're considering fit in the following framework:

\begin{Definition}\label{defConv}A couple of financial assets $(X,Y)$ are said to have converging prices if $\inf\{t\in\mathbb R_+ \;|\; X_t=Y_t\}$ is a bounded $\bG$-stopping time.
\end{Definition}

When $X$ and $Y$ are converging prices, we shall consider that the insider's horizon is a given point of convergence of the two prices,
i.e., $T$ is such that
\[
\xi:=X_T=Y_T
\] and such that $T$  is a bounded $\bG$-stopping time. One can take $T= \inf\{t\in\mathbb R_+\;|\; X_t=Y_t\}$, but such a restriction is not necessary.  In some situations the $\bG$-stopping time $T$ can be chosen as the maturity of the assets, when the cash flow $\xi$ is paid to the investors that have long positions either in the asset  $X$ or $Y$. In this case $T$ should be an $\bF^X$ and an $\bF^Y$-stopping time (i.e., cash flows are always observable by holders of long positions in the corresponding assets). Another interesting  situation  is when  $T$ is only observed by the insider, hence $T$ is neither an $\bF^X$  nor  an $\bF^Y$-stopping time. Either of the two interpretations are possible here, i.e., we do not require $T$ to be more than a bounded $\bG$-stopping time, but remaining fixed through the analysis.

Typically, two converging prices  $X$ and $Y$ may follow different paths if the different  investors, (namely the investors  active in the market for $X$ versus  the market for $Y$)  have access to different information, in which case  $\bF^X$ and $\bF^Y$ differ, and/or they have different risk attitudes. What we mean by different risk attitudes is the property that  the restriction of $\bQ^X$ to $\sigma(\xi)$ does not coincide with the restriction of $\bQ^Y$ to $\sigma(\xi)$.

\bigskip

Our aim is to analyse the no arbitrage property (NFLVR-S) from the insider's perspective, i.e., when there are prohibitions for the insider to sell short the assets $X$ and $Y$.  In other words,  we consider that the investor's strategies involve the following  positions:  long or short in cash ($\pi^C$) and only long positions in $X$
and $Y$ ($\pi^X$ and $\pi^Y$), consequently the value of the investor's portfolio writes:
\be
 V^\pi_t:=\pi^C_t+\pi^X _tX_t+\pi^Y_t Y_t,
\ee   and, when self financing, we have $dV^\pi_t:= \pi^X _tdX_t+\pi^Y_t dY_t$.
As usual, we impose some admissibility conditions for strategies under (NFLVR-S) in this framework. We refer to Pulido \cite{Pulido14} for more details.

\begin{Definition}\label{DefTradingStrategy}
A  trading strategy is a $\bG$-predictable process $\pi = (\pi^C,\pi^X,\pi^Y)$. A trading strategy $\pi$ is called an admissible trading strategy under short sales prohibitions for $X$ and $Y$ if:
\begin{itemize}
\item[(i)] $\pi^X\in L(X)$ and $\pi^Y\in L(Y)$ (i.e., $\pi^X$ is  integrable   with respect to the semimartingale $X$, $\pi^Y$ is integrable  with respect to the semimartingale  $Y$).
\item[(ii)] The process $V^{\pi}$ is bounded from below.
\item[(iii)] $\pi^X\geq 0$ and $\pi^Y\geq 0$.
\end{itemize}
We denote by $\mathcal A$ the set of admissible trading strategies under short sales restrictions for $X$ and $Y$.
\end{Definition}

 In the definition above, the price processes $X$ and $Y$ need to be $\bG$-semimartingales; this question is examined in the next section.

We now define the following sets:
\[
\mathcal K:=\{V^\pi_T,\pi\in\mathcal A\}\quad \mathcal C:=\left(\mathcal K-L^0_+(\bP)\right)\cap L^\infty(\bP).
\]where $L^0_+(\bP)$ is the space of equivalence classes of nonnegative finite random variables, and $L^\infty(\bP)$ is the space of $\bP$-essentially bounded random variables.
No Free Lunch with Vanishing Risk under short sales prohibition (NFLVR-S) is defined as follows: (NFLVR-S) holds if $\bar{\mathcal C} \cap L^0_+(\bP) =\{0\}$, where $\bar{\mathcal C}$ is the closure of $\mathcal C$ with respect with the $\|\cdot\|_\infty$ norm in $L^\infty(\bP)$.
\begin{Theorem}\cite{Pulido14}\label{NFLVR-S}
 (NFLVR-S) holds if and only if there exists a probability measure $\widetilde P $ such that  $\widetilde \bP\sim\bP$ such that the processes $X$ and $Y$ are $(\bG,\widetilde \bP)$-supermartingales. Such a probability measure is called a supermartingale measure.
\end{Theorem}

Because the condition of  no arbitrages in the form of  (NFLVR-S)
is equivalent to the existence of a supermartingale measure  for
the couple $(X,Y)$ in the filtration $\bG$, our aim is to shed light on the properties of
processes $X$ and $Y$ when considered as stochastic processes in
the larger filtration $\bG$, under the requirement that there
exists a probability measure $\widetilde \bP$ such that $\widetilde \bP\sim\bP$ such that the processes $X$ and $Y$ are $(\bG,\widetilde \bP)$-supermartingales.

\section{Structure conditions under (NFLVR-S)}\label{Section3}

We aim to clarify the properties of processes $X$ and $Y$ that admit an equivalent supermartingale measure. Structure conditions for asset prices first appeared in the setting of no arbitrage without short selling constraints, i.e.,  derived from imposing the existence of a strict martingale density. We refer to F\"ollmer and Schweizer \cite{FoelSchw91}, Ansel and Stricker \cite{AnselStick92}, Schweizer \cite{Schw95} for more details.

We shall carry our analysis in the filtration $\bG$ and the interval $[0,T]$, i.e.,  the insider's information set and the insider's investing horizon.  Note however that the stochastic processes $X$ and $Y$ are defined on infinite time horizon and the conditions (NA-X) and (NA-Y) are also supposed to hold on an infinite time horizon.

To begin with, let us introduce some notations that are going to be used in the remaining of the paper:

\begin{notation}
\begin{itemize}
\item[(i)]We write $\langle
Z\rangle$ for the sharp bracket of a semimartingale $Z$ under the
measure $\bP$ and in the filtration $\bG$. Whenever the underlying
filtration we are considering is not $\bG$ and/or the probability
is not $\bP$ we shall use explicit notations:  for instance
$\langle Z\rangle ^{(\bF,\bQ)}$ is the sharp bracket under a
measure $\bQ$ and in a filtration $\bF$ (implicitly, $Z$ needs to
be an $\bF$-semimartingale).
\item[(ii)] The expectation operator under the probability $\bP$ is written $\E$; whenever the probability measure is a different one, we shall use explicit notations, i.e., $\E^\bQ$ is the expectation under the probability measure $\bQ$.
\item[(iii)] $\mathcal
P(\bF)$ is the class of $\bF$-predictable processes, where $\bF$ is a
given filtration.
\item[(iv)] $\mathcal S(M)$ is the stable subset of $(\bG,\bP)$-local martingales generated by $M$, where $M$ is a  $(\bG,\bP)$-locally square integrable martingale ; $\mathcal S(M)^ \bot$ is the  set of $(\bG,\bP)$-locally square integrable martingales that are strongly orthogonal to $M$.
\item[(v)] $\mathcal E (Z)$ denotes the Dol\'eans-Dade exponential of a semimartingale $Z$.
\end{itemize}
\end{notation}
The following result is a more precise formulation of Theorem \ref{NFLVR-S} in the particular case of converging prices:

\bl
Suppose that $(X,Y)$ are converging prices. Then, the prices $(X,Y)$ satisfy (NFLVR-S) if and only if there exists a probability measure $\widetilde \bP$ such that $\widetilde \bP\sim\bP$ and:
\begin{align*}
X &=\widetilde M +\widetilde Z^X \\
Y &=\widetilde M +\widetilde Z^Y,
\end{align*}where $\widetilde M_t:=\E^{\widetilde \bP}[\xi |\G_t]$ and $\widetilde Z^X$ and $\widetilde Z^Y$ are two $(\bG,\widetilde \bP)$-potentials (i.e., are positive supermartingales satisfying $\widetilde Z^X_T=\widetilde Z^Y_T=0$).
\el

\proof (NFLVR-S) holds if and only if a supermartingale measure
$\widetilde \bP$ exists. But then $X$ and $Y$ are uniformly
integrable $\widetilde \bP$-supermartingales and the expressions
follow from the Riesz decomposition  and the terminal
condition $X_T=Y_T$. For more details, see \cite{Meyer}-VI-11or, alternatively, \cite{cdellacherie} T12 p. 97.\finproof

Now, we investigate the structure of the two price processes under the reference probability
$\bP$, which is arbitrarily chosen.

\bp\label{ThmSemi}
Assume the prices $(X,Y)$ satisfy (NFLVR-S).  If $\E[X_T]<\infty$ and $\E[Y_T]<\infty$ , then $(X,Y)$ are $(\bG,\bP)$-special semimartingales on the interval $[0,T]$.
 \ep

\proof If $(X,Y)$ satisfy (NFLVR-S), then there exists a
probability measure equivalent to $\bP$, say $\widetilde \bP$,
such that  $(X,Y)$ are $(\bG,\widetilde \bP)$-supermartingales.
The set of semimartingales being stable under equivalent changes
of the probability measure (Girsanov-Meyer theorem), it follows
that $(X,Y)$ are $(\bG,\bP)$-semimartingales. We now show that $X$ is a special  $(\bG,\bP)$-semimartingale; the reasoning for $Y$ is identical.

First we prove the property in the filtration $\bF^X$ and then in the larger filtration $\bG$.
 We  denote $Z_k:=\frac{d\bP}{d\bQ^X}|_{\F^X_k}$ (with $k$
such that  $T\leq k$ $a.s.$) and $Z_t= \E(Z_k \vert \F^X_t)$. We have:
\[
\E^{\bQ^X}\left[[X,Z]_T\right]\leq \E^{\bQ^X}\left[X_T Z_T\right ]-X_0+\E^{\bQ^X}\left[N^*_k\right ] =\E^{\bP}[X_T]-X_0+ \E^{\bQ^X}\left[N^*_k\right ],
\]
with $N^*_k:=\sup_{s\in[0,k]}|N_s|$ where $N=-\int Z_-dX-\int
X_-dZ$. The process $N$ is a $(\bF^X,\bQ^X)$-local martingale ($X$ and $Z$ being
$(\bF^X,\bQ^X)$- martingales). Therefore $N^*$ is locally
integrable under $\bQ^X$ and so is $[X,Z]$, due to the inequality
above.
 Hence the $(\bF^X,\bQ^X)$-predictable bracket for $X$ and $Z$, $\langle X,Z\rangle^{(\bF^X,\bQ^X)}$ exists. It follows by   Girsanov's theorem that:
 \[
X_t = m^\bP_t +\int_0^t\frac{1}{Z_{s-}}d\langle X,Z\rangle^{(\bF^X,\bQ^X)}_s
 \] where $m^\bP$ is an $(\bF^X,\bP)$-local martingale. Hence, the  process  $X$ is  a special semimartingale in $\bF^X$. We examine the situation in the larger filtration $\bG$ and we remark that it is sufficient to show that $m^\bP$ is a special $(\bG,\bP)$-semimartingale.

The process $m^{\bP}$ is a $(\bG,\bP)$-semimartingale ($X$ being one). Moreover,  $\sup_{s\leq t}|m^\bP_s|$ is $(\bF^X,\bP)$-locally integrable ($m^\bP$ is a $(\bF^X,\bP)$-local martingale, hence we use Theorem 34 p.130 in \cite{protter}) and therefore it is also $(\bG,\bP)$-locally integrable ($\bF^X$-stopping times are $\bG$-stopping times). This in turn implies that $m^\bP$ is a $(\bG,\bP)$-special semimartingale (Theorem 33 p.130 in \cite{protter}) which proves the result.

\finproof

\bethe\label{ThmGenRepr} Assume that  $(X,Y)$ satisfy
(NFLVR-S). If  $\E[X_T]<\infty$, then there exist $J^X$ and $w^X$ all being
in $ \mathcal P(\bG)$ and a $(\bG,\bP)$-local martingale  $M^X$ with $M^X_0=0$, such that for any $t\leq T$:
\begin{align}\label{Xrepr}
X_t&=X_0+J^X_t+\int_0^t w^X_ud\langle M^X \rangle_u+M^X_t\,.
\end{align}
If $X$ is $\bF^X$-predictable, the process $J^X$  is null.
In general, the process  $J^X$  satisfies $J^X_0 =0$, is   decreasing, and $dJ^X$ is singular with respect to $d\langle M^X\rangle$.
\eethe

\proof
 In view of Proposition \ref{ThmSemi}, there exists a $(\bG,\bP)$-local martingale $M^X$ and a finite variation, $\bG$-predictable process $V^X$, such that:
\begin{align*}\label{Yrepr}
X_t&=X_0+V^X_t+M^X_t.
\end{align*}
We can write $V^X_t=\int_0^t w^X_ud\langle M^X\rangle_u+J^X_t$,
where $dJ^X$ is a signed measure that is singular with respect to
$d\langle M^X\rangle$ (i.e., the Lebesgue decomposition of
$dV^X$ with respect to $d\langle M^X\rangle$; see Proposition \ref{AnnexA3} in Appendix \ref{A}).

To show that $J^X$ is a decreasing process we use
Girsanov's  theorem and Theorem \ref{diffmeasures} in Appendix \ref{A}. More precisely, let $\tilde \bP$ be an equivalent $\bG$-supermartingale measure for $X$. By Girsanov's theorem the decomposition of $X$ is given by: $X =X_0+(J^X  +\tilde D^X )+\tilde M^X $, where $\tilde M^X$ is a $(\bG,\tilde\bP)$-martingale and:
\begin{itemize}
\item[(i)]  $d\tilde D^X\ll d\langle M^X\rangle$. Hence  $d\tilde D^X$ and  $dJ^X$ are orthogonal.
\item[(ii)] the process $ J^X  +\tilde D^X  $ is decreasing.
\end{itemize}The two above points imply that both $J^X$ and $\tilde D^X$ are decreasing (Theorem \ref{diffmeasures} (b)), hence the statement.

Let us suppose that $X$ is $\bF^X$-predictable, hence continuous
(because $X$ is also an $(\bF^X,\bQ^X)$-local martingale). Then, $[X]=
\langle M^X\rangle$ and the process $X$ has the same constancy
intervals as $\langle M^X\rangle$. It follows that $J^X\equiv
J^X_0$, i.e., is constantly null.
\finproof

 We emphasise that from Theorem \ref{ThmGenRepr}, (NFLVR-S) and $\E[Y_T]<\infty$  imply a decomposition for $Y$ (with obvious notations):
\begin{equation}\label{Yrepr}
Y_t=Y_0+J^Y_t+\int_0^t w^Y_ud\langle M^Y \rangle_u+M^Y_t. 
\end{equation}

We consider below two examples of converging prices.

\begin{exam} Let $B^1$ and $B^2$ be two independent
$\bP$-Brownian motions with respective natural filtrations $\bF^1$
and $\bF^2$; consider that $\theta^1$ is a $\bF^1$ stopping
time and $\theta^2$ is a $\bF^2$ stopping time (hence, they
are predictable), both considered to have absolutely continuous distribution functions denoted $C^1$ and $C^2$, and satisfying $C^1(T)<1,C^2(T)<1$.

The following payoff is scheduled at a fixed maturity date $T$:
\[
\xi=\ind_{\{\theta_1> T\}}+\ind_{\{\theta_2\leq T\}}.
\]
We consider the following distinct information sets:
\begin{align*}
\G^1_t:= &\F^2_t\vee\sigma(t\wedge \theta^1),\\
\G^2_t:= &\F^1_t\vee\sigma(t\wedge \theta^2).
\end{align*} and we assume that the corresponding prices are $X_t=\bP(\xi|\G^1_t)$ and $Y_t=\bP(\xi|\G^2_t)$.

We have the following $\bG^1$ martingales, $t\leq T$: 
\[
\bP(\theta^1>T|\G^1_t)=\ind_{\{\theta^1>t\}}\frac{\bP(\theta^1>T)}{\bP(\theta^1>t)}=\ind_{\{\theta^1>t\}}\frac{1-C^1(T)}{1-C^1(t)}
\] (see Proposition 1 in \cite{ejy}), and:
\[ 
\bP(\theta^2\leq T|\G^1_t)=\bP(\theta^2\leq T|\F^2_t),
\] (as $\theta^2$ is independent from $\theta^1$) i.e., the last process is a Brownian martingale. 
We deduce that the $\bG^1$ adapted price for the claim $\xi$ decomposes as follows:
\begin{align*}
X_t=& X_0-\int_{0}^t \frac{1-C^1(T)}{1-C^1(s)}d\ind_{\{\theta_1\leq s\}} +\int_0^{t\wedge \theta^1}(1-C^1(T))d(1-C^1(s))^{-1}  +M^X_t,
\end{align*}
where $M^X=\bP(\theta^2\leq T|\F^2_\cdot)-\bP(\theta^2\leq T)$.

Similar arguments lead to the following $\bG^2$ adapted price:
\begin{align*}
Y_t=&Y_0+\int_{0}^t \frac{1-C^2(T)}{1-C^2(s)}d\ind_{\{\theta_2\leq s\}} -\int_0^{t\wedge \theta^2}(1-C^2(T))d(1-C^2(s))^{-1}  +M^Y_t,
\end{align*}
with $M^Y=\bP(\theta^1> T|\F^1_\cdot)-\bP(\theta^1> T)$.

 One can check that: $\bF^X=\bG^1$,  $\bF^Y=\bG^2$, while the insider filtration is $\bG=\bF^1\vee\bF^2$ (i.e., the natural filtration  of $(B^1,B^2)$). It follows that $M^X$ and $M^Y$ are also $\bG$ martingales. They are Brownian martingales, from the discussion above. Therefore $\langle M^X\rangle$ and $\langle M^Y\rangle$ are absolutely continuous with respect to the Lebesgue measure.
 We deduce that $X$  decomposes as in  (\ref{Xrepr}) and $Y$ as in (\ref{Yrepr}), with the processes
\begin{align*}
J^X:&=\left(-\int_0^t \frac{1-C^1(T)}{1-C^1(s)}d\ind_{\{\theta_1\leq s\}}\right) \\
J^Y:&=\left(\int_0^t \frac{1-C^2(T)}{1-C^2(s)}d\ind_{\{\theta_2\leq s\}}\right)
\end{align*} being $\bG$-predictable.  Because $J^Y$ is an increasing process, we conclude by Theorem \ref{ThmGenRepr} that the price process $Y$ does not respect (NFLVR-S) for the insider. Indeed, $\theta^2$ being a predictable $\bG$ stopping time, the insider can buy the asset $Y$ at time $\theta^2_-$ and resell it at time $\theta^2$ thus making an arbitrage profit of one monetary unit.
\end{exam}

\begin{exam}
Let us consider the hitting time by a Brownian motion $B$ of a positive random variable $D$  independent from the Brownian motion:
\[
T^D=\inf\{t\geq 0\;|\; B_t\geq D\}.
\]
In the filtration    $\bF$ given by $\F_t:=\sigma(T^D\wedge s,s\leq t)$ we have that
$T^D$ is a totally inaccessible $\bF$-stopping  time with corresponding $\bF$-intensity process:
\[
c(t)=\frac{\ind_{\{T^D>t \}}}{\bP(T^D>t)}\int_0^\infty f_x(t)dF_D(x),
\]
where  $F_D(x)$ is the distribution function of $D$   and $f_x(t)$ is the density function of the hitting time $T^x$.  We denote $H_t: =\ind_{\{T^D\leq t\}}- \int_0^{t}c(s)ds$ which is an $\bF$-martingale.

Let us assume that the price process $X$ is given by the positive local martingale  $X=X_0\mathcal E(-H)$, ithat is, it satisfies:
\[
X_t =X_0-\int_0^t X_{s_-} dH_s.
\]
One can notice  that $\bF^X=\bF$. For simplicity we do not introduce the second asset $Y$ and we  rather concentrate on the dynamics of $X$ in the larger filtration
$\bG$ given by $\G_t:=\F^X_t\vee \sigma(B_s, s\leq t)$.

 We denote $\Lambda^\bG$ the $\bG$-compensator of $T^D$, so that the process: $H^\bG_t:=\ind_{\{T^D\leq t\}}-\Lambda^\bG_t$ is a martingale.  It can be shown  (using \cite{ejy} and the fact that $\sigma(B_s, s\leq t)$ is immersed in $\bG$),  that $\Lambda^\bG$ is absolutely continuous with respect to
the measure generated by the running supremum of the Brownian motion:
\[
\Lambda^\bG_t=\int_0^{t\wedge T^D} \frac{d\bP(T^D>s|\F^B_s)}{\bP(T^D>s|\F^B_s)}=-\int_0^{t\wedge T^D} \frac{dF_D(S_s)}{1-F_D(S_s)}= -\ln F_D(S_{t\wedge T^D}) ,
\]
where $\bF^B$ is the Brownian filtration and $S$ is the running supremum of $B$.
The $\bG$ decomposition of $X$ writes, using that $H=H^\bG-\int c(s)ds+\Lambda ^\bG$:
\[
X_t =X_0+\left(\int_0^tX_sc(s)ds-\int_0^t X_sd\Lambda^\bG_s\right)- \int_0^t
X_{s-} dH^\bG_s.
\]
 Using Theorem \ref{ThmGenRepr} we identify $M^X_t=- \int_0^t
X_{s-} dH^\bG_s$. We have shown  above that $d\langle M^X\rangle$  is  absolutely continuous with respect to $dS$. Therefore,
  $J^X_t=\int_0^tc(s)ds$, as the Lebesgue measure is orthogonal with respect to the $dS$.     Because $J^X$ is increasing, from Theorem \ref{ThmGenRepr} we conclude that
 there are arbitrage opportunities, in the sense that
(NFLVR-S) fails. Here again, an arbitrage strategy is easy to
implement by the $\bG$-informed investor: buy the asset $X$ at any
time before $T^D$ when the Brownian motion is strictly below its
running maximum and sell it any time before it reaches its maximum
level again.    On these
intervals, the price process $X$ is strictly increasing; the arbitrage
strategy described performs a strictly positive profit proportional to the
holding period of the asset $X$.
\end{exam}

\section{A Result on the Existence of a Supermartingale Measure}\label{sec::MainResult}

In this section we investigate the existence of a specific
$\bG$-supermartingale measure for two price processes $X$ and $Y$,
that we shall call   fundamental supermartingale measure for $(X,Y)$. This  object will play
an important role, as systematic arbitrage opportunities occur
when this supermartingale measure cannot be constructed.

 It is convenient to take as underlying probability measure a specific $\bG$-supermartingale measure for $X$, that for simplicity we still call $\bP$, so that
in the filtered probability space  $(\Omega, \G,\bG,\bP)$, the two assets  have the following representations:
\begin{align}\label{PX}
X&=X_0+J^X+M^X,\\\label{PY}
Y&=Y_0+V^Y+M^Y,
\end{align}
with $M^X$ and $M^Y$ being $(\bG,\bP)$-local martingales that are locally square integrable with $M^X_0=M^Y_0=0$ and such that the process $V^Y$ is a finite variation, $\bG$-predictable process.
The process $J^X$ is considered to be decreasing and the measure $dJ^X$ is orthogonal to $d\langle M^X \rangle$.

Notice that the decomposition in (\ref{PX})  is not the same as
in (\ref{Xrepr}). In the previous section $\bP$ was an arbitrary probability measure (equivalent to $\bQ^X$ and $\bQ^Y$). In this section $\bP$ is a particular equivalent supermartingale measure for $X$, such that $X$ decomposes precisely as in (\ref{PX}). The existence of such a supermartingale measure for $X$ -that here is assumed- is a first step to the construction of the
fundamental supermartingale measure for the couple $(X,Y)$.

We decompose the martingale $M^Y$ as:
\[
M^Y =M^1 +M^2
\]
with $M^1\in \mathcal S(M^X)$ and $M^2\in \mathcal S(M^X)^\bot$ so that we can write $M^1$ as:
\be\label{M1}
M^1_t=\int_0^t h_u dM^X_u,
\ee
for some process $h\in\mathcal P(\bG)$ and assumed to  have right-continuous sample path.

\bigskip

We shall need the following additional decompositions:

\begin{itemize}

\item The predictable, finite variation part of $Y$ stated in  (\ref{PY})  decomposes as:
\be\label{V=A-a}
V^Y =A -a ,
\ee
where $A$ and $a$ are increasing processes which do not increase on the
same sets (that is $dA$ and $da$ are orthogonal measures)\footnote{In order to preserve the compatibility with  the decomposition result in Theorem \ref{ThmGenRepr}, $dA$ is assumed absolutely continuous with respect to $d\langle M^Y\rangle$. This property will solely be used for constructing an arbitrage portfolio in Lemma \ref{Arbitrage}. }\label{foot}, and $a_0=A_0=0$.

\item Moreover, the process $A$ can always (and uniquely) be written as sum of
two other increasing processes:
\[
A =A^{1} +A^{2} ,
\] where $dA^{1}  \ll h^+ d\langle M^X \rangle $ and $dA^{2}  \bot h^+ d\langle M^X \rangle $.  Therefore, there exist  $a^1\geq 0$ such that
\[
A^1_t=  \int_ 0^t    a^1_u h^+_u d\langle M^X\rangle _u
\] (the non negativity of $a^1$ comes from the fact that $A^1$ is increasing)  and
  $\tilde a^{1} $  so that $\tilde a^1 =\tilde a^1  \ind_{ {\{h> 0 \}}}= \frac{a^1}{h}  \ind_{ {\{h> 0 \}}}\geq 0$  and
\[
A^1_t=\int_0^t \tilde a^1_ud\langle M^1\rangle _u = \int_ 0^t \tilde a^1_u (h_u)^2  d\langle M^X\rangle _u.
\]

\end{itemize}

\bigskip

We now state our main result of this section:
\bethe\label{MainThm}
 Assume that $\tilde a^{1}\Delta M^1<1$ holds almost surely.  We consider the following
conditions:
\begin{itemize}
\item[(C1)] $dA^{2}  \ll d\langle M^2 \rangle  $. We denote $\tilde a^{2} $ the density of $dA^{2} $ with respect to $d\langle M^2\rangle $.
\item[(C2)]  $\tilde a^{2}\Delta M^2<1$.

\item[(C3)] $\E\left [D^*_T\right ]=1$, where:
\be
D^*_t:= \mathcal  E_t\left( -\int_0^\cdot \tilde a^1_sdM^1_s \right)
\mathcal E_t\left( -\int_0^\cdot \tilde a^{2}_sdM^2_s\right),\quad t\in[0,T].
\ee
\end{itemize}
If (C1)-(C3) are satisfied, the price processes $(X,Y)$ satisfy (NFLVR-S). Additionally,  the probability measure $\bP^*$ defined as:
\be\label{MSM}
\frac{d\bP^*}{d\bP}\Big|_{\G_t}: =D^*_t,\quad t\in[0,T].
\ee
is a supermartingale measure for $(X,Y)$ that we call the fundamental supermartingale measure  for $(X,Y)$.

Conversely, if the price processes $(X,Y)$ satisfy (NFLVR-S), then (C1) and (C2) hold true, so that the process $D^*$ is a strictly positive local martingale.
\eethe

Before proving the theorem, let us give some simple examples in order to illustrate the many processes involved, in particular the different decompositions of the process $V^Y$.  Note that we do not consider below that $X$ and $Y$  are converging prices; examples with converging prices are provided in Section \ref{sec::Examples}.

\begin{exam}
Suppose that $B^1$ and $B^2$ are two independent Brownian motions and
\begin{align*}
X_t&=X_0+B^1_{t\wedge \theta}\quad\text{ with }\theta:=\inf\{t\in[0,T], X_t=0\}, T\text{ fixed}\\
Y_t&=Y_0+\int_0^t F_sds+\int_0^tH_s dB^1_s+\int_0^tG_sdB^2_s,  t\in[0,T],
\end{align*}
with $H$ and $G$ some predictable processes, that for simplicity we assume bounded. Let us identify the key processes introduced previously in this section.

We have $h_t=H_t$ and: $\langle M^1 \rangle _t=\int_0^{t\wedge \theta} (H_s)^2ds$; $\langle M^2 \rangle _t=\int_0^t \left[(H_s)^2\ind_{\{\theta\leq s\}} +(G_s)^2 \right]ds$.  Moreover,
 the process $A_t= \int_0^t (F_s)^+ ds $ decomposes $A=A^1+A^2$ with:
\begin{align*}
A^1_t&=\int_0^{t\wedge \theta} \ind_{\{H_s>0\}}(F_s) ^+ds=\int_0^t\tilde a^1_sd\langle M^1\rangle_s, & \mbox{where}\quad \tilde a^1_t &=\frac{ \ind_{\{H_t>0\}}(F_t)^+}{(H_t)^2}\\
A^2_t&=\int_0^t \ind_{\{H_s\leq 0\}\cup \{\theta\leq s\}}(F_s)^+ds.
\end{align*}The existence of the density process $\tilde a^2$ is not guaranteed. The absolute continuity condition (C1) in the Theorem \ref{MainThm} becomes:
The process  $G$ is non null on the set:
\[
\{ (t,\omega) |\theta(\omega)>t, H_t(\omega)\leq 0,F_t(\omega)>0\} \cup \{ (t,\omega) |\theta(\omega)\leq t, H_t(\omega)= 0,F_t(\omega)>0\}.
\]
When this is the case, we have $A^2_t=\int_0^t\tilde a^2_sd\langle M^2\rangle_s$ with
\[
 \tilde a^2_t= \ind_{\{\theta> t\}}\frac{\ind_{\{H_t\leq 0\}}(F_t)^+}{(G_t)^2}+\ind_{\{\theta \leq t\}}\frac{(F_t)^+}{(H_t)^2+(G_t)^2}
\]
 and the following process
\[
D^*:=\mathcal E \left(-\int_0^{\theta\wedge \cdot} \ind_{\{H_s> 0\}}\frac{(F_s)^+}{H_s} dB^1_s \right)\mathcal E \left(-\int_0^\cdot \ind_{\{h_s \leq 0\}}f(B^1_s)^+ dB^2_s \right).
\] is the candidate for the density of the fundamental supermartingale measure.
The theorem then states that there exists a super-martingale measure if (the other conditions being fulfilled) $\E\left[D^*_T\right]=1$.

\end{exam}

\begin{exam}Another simple example is the one where $M^2\equiv 0$. In this case the theorem simply says that $A$ should not increase on the sets where $d\langle X,Y\rangle< 0$, otherwise (NFLVR-S) does not hold.  See also Subsection \ref{Example1}.
\end{exam}

\begin{exam} If the process  $ \langle X,Y\rangle$ is strictly increasing, then $A^2\equiv 0$ and only the condition (C3):  $\E\left[\mathcal  E_T\left( -\int_0^\cdot \tilde a^1_sdM^1_s \right)\right]=1$ needs to be checked. However, if this not fulfilled, we cannot  in general conclude to absence of (NFLVR-S) as (C3) is not a necessary condition.
 \end{exam}

\bigskip

\proof \textit{(Proof of the Theorem \ref{MainThm})}

"$\Leftarrow$" Condition (C1) ensures the existence of a process
$\tilde a^2$, such that:
\[
A^2_t=\int_0^t \tilde a^2_ud\langle M^2\rangle _u
\]and $\tilde a^2$ is nonnegative, due to the increasing property
of $A^2$. Condition (C3), implies that the process $D$ is a
martingale, while condition (C2) ensures that it is strictly
positive (indeed, if a semimartingale $H$ satisfies $\Delta H>-1$,
then the stochastic exponential process  $\mathcal E (H)$   is
strictly positive. If we take $H_t:=-\int_0^t \tilde a^{2}_s
dM^2_s$, we get that $\Delta H=-\tilde a^{2}\Delta M^2$; the corresponding
 condition for $\tilde a^{1}$ was already assumed to hold).

We define:
\[
\frac{d\bP^*}{d\bP}\Big|_{\G_T}:=D^*_T
\]
which is indeed an equivalent probability measure. It is easy to
check that it is a supermartingale measure: indeed, under $\bP^*$,
$$dX_t=dJ_t^X- \tilde a^1_t h_t d \langle M^X\rangle _t+ dm^*_t =dJ^X_t-\tilde a^1_t (h_t)^+ d \langle M^X\rangle _t+  dm^*_t
$$  where $m^*$ is a
$\bP^*$-martingale.   The processes $J^X$ and $-\int  \tilde a^1_s(h)^+_s d \langle M^X\rangle_s$ being decreasing, $X$ is a supermartingale under $\bP^*$.  Also:
$$dY_t= dV^Y_t+dM^Y_t= dA^1_t+dA^2_t-da_t  +dM^* _t- \tilde a^1_t (h_t)^2 d \langle M^X\rangle _t-  \tilde a^2_t d \langle M^2\rangle _t=    dM^*_t -da_t$$
where $M^*$ is a $\bP^*$-martingale.
\bigskip

"$\Rightarrow$" We assume that there exists an equivalent
supermartingale measure, that we denote $\widetilde \bP$. Without loss of
generality, the density process has the representation
\be\label{measuretilde}
\frac{d\widetilde \bP}{d\bP}\Big |_{\G_t}=\mathcal E(-L)_t,
\ee where $L$ can be decomposed as:
\[
L_t=\int_0^t\ell^1_udM^1_u+\int_0^t\ell^2_udM^2_u+U_t.
\]with $U$ a local martingale orthogonal to both $M^1$ and $M^2$.

 We use the notation $\{dA\neq 0\}$ for the support of the measure $dA (\omega)$.

The processes $X$ and $Y$ are $\widetilde \bP$-supermartingales; therefore we need to have
simultaneously:

\begin{enumerate}
\item[(i)] $(\langle M^X,L\rangle_t, t\in [0,T])$ is an increasing process;
\item[(ii)] $\left (\int_0^t\ind_{\{dA\neq 0\}}d\langle
M^Y,L\rangle_u - A_t,t\in[0,T]\right )$ is a decreasing process.
\end{enumerate}

Condition (i) is obtained as follows. The process $X$ is a  $\widetilde \bP$-supermartingale if and only if $J^X  -\langle M^X, L\rangle$ is a decreasing process. But $J^X$ is decreasing and $d J^X$ is orthogonal to  $d \langle M^X\rangle$, therefore the condition (i) appears as necessary and sufficient for $X$ to be a   $\widetilde \bP$-supermartingale.

Also, some clarifications concerning the condition (ii) above. The process $Y$ is a  $\widetilde \bP$-supermartingale if and only if $ V^Y -\langle M^Y,L\rangle $  is a decreasing process. But $ V^Y -\langle M^Y,L\rangle $ is decreasing if and only if the two processes $(A_t-\int_0^t\ind_{\{dA\neq 0\}}d\langle M^Y,L\rangle_u,t\in[0,T])$  and $(-a_t-\int_0^t\ind_{\{da\neq 0\}}d\langle M^Y,L\rangle_u,t\in[0,T])$ are decreasing. However, the last condition is not going to be exploited here.

From condition (i) above, we obtain that necessarily the process  $ h\ell^1 $ is positive (i.e., nonegative). In particular on the set $\{h<0\}$ the process $\ell^1$ has negative or null values only.

Let us now analyze condition (ii). For simplicity, we denote:
$\tilde \ell ^1_t:= \ell^1_t\ind_{\{dA\neq 0\}}$ and $\tilde \ell^2_t:=
\ell^2_t\ind_{\{dA\neq 0\}}$. From the above observation regarding $\ell^1$, the process $\tilde \ell^1$ satisfies:
\begin{equation}\label{tildel1}
\tilde \ell^1\ind_{\{h<0\}}\leq 0,
\end{equation}

We recall that the process $ A $
decomposes as $\int_0^t \tilde a^1_ud\langle M^1\rangle_u +A^2_t$,
with $\tilde a^1$ satisfying $\tilde a^1 =\tilde a^1 \ind_{\{h > 0\}}$ and hence:
\begin{align*}
&\left(\int_0^t\ind_{\{dA\neq 0\}}d\langle M^Y,L\rangle_u-A_t\right)=\\
&=\int_0^t (\tilde \ell^1_u-\tilde a^1_u)\ind_{\{h_u> 0\}}d\langle M^1\rangle_u+\int_0^t \tilde
\ell^2_ud\langle M^2\rangle_u-\left (A^2_t- \int_0^t \tilde \ell^1_u\ind_{\{h_u\leq 0\}}d\langle M^1\rangle_u \right).
\end{align*}
The process above should be increasing. Because both
processes $  A^2 $ and $ -\int_0^ \cdot \tilde
\ell ^1_u\ind_{\{h_u\leq 0\}}d\langle M^1\rangle_u = -\int_0^ \cdot \tilde
\ell ^1_u\ind_{\{h_u< 0\}}d\langle M^1\rangle_u $ are increasing (see (\ref{tildel1})) and they
do not increase (i.e., they stay constant) on the set $\{h_t>
0\}$, it follows that the process:
\begin{equation}\label{decr}
 \int_0^\cdot \ind_{\{h_u\leq 0\}} \tilde \ell^2_ud\langle M^2\rangle_u-C 
\end{equation}
 needs to  be increasing, where $C_t:=A^2_t-\int_0^t \tilde \ell^1_u\ind_{\{h_u\leq 0\}}d\langle M^1\rangle_u$ is increasing. It follows from Theorem \ref{diffmeasures}
 in the Appendix A that $C$ is absolutely continuous with respect
to $d\langle M^2\rangle $. Because $C$  is the sum of two increasing processes, then each term should be absolutely continuous with respect to $d\langle M^2\rangle $, that is:
\begin{equation}\label{ell1bis}
\int_0^t \tilde \ell^1_u\ind _{\{h_u\leq0\}}d\langle M^1\rangle_u=\int_0^t
\tilde \ell^1_u\ind_{\{h_u\leq0\}}e_u d\langle M^2\rangle_u
\end{equation}
for some nonnegative process $(e_t)$, and
\[
A^2_t=\int_0^t\tilde a^2_ud\langle M^2\rangle _u
\]
 for a nonnegative process $ \tilde a^2 = \tilde a^2  \ind_{\{h \leq0\}}$.
It follows that the condition (C1) in the theorem must hold. In particular, the local martingale $D^*$ exists.

It remains to show that (C2) holds as well, a property that triggers the strict positivity the local martingale $D^*$. Below we show that (C2) is a consequence of the strict positivity of the local martingale $ \mathcal E(-L) $ in (\ref{measuretilde}). We notice first that, the process in (\ref{decr}) being decreasing:
\[
(\tilde a^2_t- \tilde \ell^1_te_t-\tilde \ell^2_t)\ind_{\{h_t\leq0\}}\leq 0,
\]
and therefore:
\begin{equation}\label{ineq1}
0\leq \tilde a^2_t\leq ( \tilde \ell^1_te_t+\tilde \ell^2_t)\ind _{\{h_t\leq 0\}}\leq \tilde \ell^2_t\ind_{\{h_t\leq0\}}.
\end{equation}
To obtain the last inequality above, we use  $\tilde \ell^1e\ind _{\{h\leq 0\}}\leq0$ ($e$ being a positive process). Indeed:  $\ind_{\{h<0\}}\tilde \ell^1\leq 0$ as in (\ref{tildel1}) and the set $\{h=0\}$ we have $\ind_{\{h=0\}}d\langle M^X\rangle =0$ and therefore, using the equality (\ref{ell1bis}) the process $\ind_{\{h=0\}}\tilde \ell^1 e =0$.

As the process $ \mathcal E(-L) $ in (\ref{measuretilde})  is
strictly positive, and  from the orthogonality of $M^1$
and $M^2$, it follows that we must have:  $-\ell ^{1}\Delta M^1>-1$ and  $-\ell ^{2}\Delta M^2>-1$. In particular, the last inequality holds on the set $\{\tilde a^2  > 0\}\cap\{h \leq 0\}$ (notice that on this set we have $\ell ^2>0$, which follows from  (\ref{ineq1})). Then, the inequalities in (\ref{ineq1}) ensure that
\[
-\tilde a^{2}\Delta M^2>-1
\]

Indeed, (\ref{ineq1}) implies that $ -\tilde a^2 \geq  -\tilde \ell^2 \ind_{\{h \leq 0\}}$, hence, if $\Delta M^2 >0$, one has $ -\tilde a^2  \Delta M^2 \geq  -\tilde \ell^2 \ind_{\{h \leq0\}}\Delta M^2 \geq  -\ind_{\{h \leq 0\}}\geq -1$. If $\Delta M^2 <0$, one has $-\tilde a^2  \Delta M^2 \geq 0>-1$.
Therefore the condition (C2)  in the theorem holds as well.
\finproof

\bigskip

Theorem \ref{MainThm} emphasizes the fact that the condition (C1) is necessary for (NFLVR-S) to hold. In the remaining of this section we reveal a systematic arbitrage portfolio when (C1) fails. For this, we identify the set where the condition fails (i.e., the arbitrage set):
\[
\mathcal A:=\{(\omega, t)\in\Omega\times [0,T(\omega)] \;|\; dA^2_t(\omega)>0 \text{ and } d\langle M^2\rangle_t(\omega)=0\};
\]in other words, in $\mathcal A$ the measure $dA^2$ is not absolutely continuous with respect to $d\langle M^2\rangle$. The condition (C1) can be rewritten as: $\bP(\omega: \exists t, (\omega,t)\in \mathcal A)=0$.

We introduce the d\'ebut of $\mathcal A$:
\[
D_\mathcal A:=\inf\{t\geq 0\;|\; (t,\omega)\in \mathcal A\},
\] with the usual convention: $\inf\emptyset=\infty$.

The random time $D_A$ is a predictable $\bG$ stopping time. This can be proved as follows.  The processes $A^2$ and $\langle M^2\rangle$ are $\bG$-predictable,
 hence the set $\mathcal A$ is $\bG$-predictable. Furthermore, $A^2$ and $\langle M^2\rangle$ are right continuous, so that $[\![D_A]\!]\subset \mathcal A$. We conclude using Proposition 2.40, p. 354 in \cite{Nikeghbali06}.

The exit time from $\mathcal A$:
\[
E_\mathcal A:=\inf\{t>D_\mathcal A\;|\; (t,\omega)\notin \mathcal A\}
\]is as well a predictable stopping time (it can be also written as the d\'ebut of the set $\{(\omega, t)\in\Omega\times [\![D_\mathcal A\wedge T,T]\!] \;|\; dA^2_t(\omega)=0 \text{ or } d\langle M^2\rangle_t(\omega)>0\}$).

To construct our arbitrage portfolio we use  a trading strategy $\pi=(\pi^C,\pi^X,\pi^Y)$, where $\pi^X_t\geq 0$ represents the
quantity of asset $X$ in the portfolio at time $t$, $\pi^Y_t\geq
0$ the quantity of asset $Y$ and  $\pi^C_t\in \mathbb R$ is the
amount invested in the risk-free asset (cash)  at time $t$ to have a self financing strategy (see Definition \ref{DefTradingStrategy}).
We recall that the value of the portfolio at time $t\in[0,T]$ writes:
\be\label{Vpi}
 V^\pi_t:=\pi^C_t+\pi^X _tX_t+\pi^Y_t Y_t.
\ee
Additionally, our arbitrage portfolio will satisfy the following conditions:
\begin{itemize}
\item[(a)] it is initiated at time $D_\mathcal A$ at no cost: $V^\pi_{D_\mathcal A}=0$.
\item[(b)] at some $\bG$ stopping time $S\leq T$ the portfolio has positive value: $V^\pi_S\geq 0$ $a.s.$ with $\bP(V^\pi_S>0)>0$. In our case $S$ is any stopping time less or equal to $E_\mathcal A$.
\item[(c)] the underlying trading strategy $\pi$ is admissible in the sense of the Definition \ref{DefTradingStrategy} (some of the admissibility conditions are already implied by the previous points).
\end{itemize}

Such a portfolio is indeed the  following one: $\pi_{0}=(0,0,0)$ (that is, no initial investment), then the self financing strategy associated with
\begin{align}\label{piX}
&\pi^X_t=-h_{t}\ind_{\{t\in[\![D_\mathcal A, E_\mathcal A[\![\}} \\\label{piY}
&\pi^Y_t=\ind_{\{t\in[\![D_\mathcal A, E_\mathcal A[\![\}}.
\end{align}    The lemma below shows that the portfolio value is increasing, in particular it is bounded from below, which ensures that the underlying trading strategies are admissible, that is, (c) is satisfied.  It also proves that condition (b) holds (i.e., the portfolio is an arbitrage) as soon as we have a violation of (C1), that is: $\bP(\omega: \exists t, (\omega,t)\in \mathcal A)>0$.

\bl\label{Arbitrage} The value of a self-financing portfolio $V^\pi$  with  $\pi$ as in (\ref{piX})-(\ref{piY}) is an increasing process, and strictly increasing for $(\omega,t)\in\mathcal A$.
\el
\proof
The portfolio value is constant outside the set $\mathcal A$, therefore we only need to investigate the behaviour of the prices processes $X$ and $Y$ inside the set $\mathcal A$.

The portfolio being self-financing, we have:
\begin{align*}
dV^\pi_t&=-h_tdX_t+dY_t=\left(-h_tdJ^X_t-h_tdM^X_t\right)+\left(dV^Y_t+h_tdM^X_t +dM^2_t\right)\\
&=-h_tdJ^X_t+dA^2_t+dM^2_t.
\end{align*}The last equality appears as a consequence of  the fact that in $\mathcal A$ we have $dA^2>0$ so that $da=dA^1=0$ and $dV^Y=dA^2$.

We recall the following properties:  $dA^2$ is absolutely continuous with respect to $d\langle M^Y\rangle$ (consequence of the fact that $dA$ is absolutely continuous with respect to $d\langle M^Y\rangle$, see footnote page 9);  and inside $\mathcal A$ we have $dA^2$ is orthogonal to $d\langle M^2\rangle$. It follows that inside $\mathcal A$, $dA^2$ is absolutely continuous with respect  to $d\langle M^1\rangle$ and hence also with respect to $\langle M^X\rangle$. On the other hand, $dJ^X$ is orthogonal to $ d\langle M^X\rangle$. It follows that in $\mathcal A$ the process $J^X$ is constant, that is: $dJ^X\equiv 0$. From this, we deduce that the dynamics of the portfolio's value can be rewritten:
\begin{align*}
dV^\pi_t&=dA^2_t+dM^2_t\text{ for }(\omega, t)\in\mathcal A.
\end{align*}
We now notice that inside $\mathcal A$ we have $d\langle M^2\rangle \equiv 0$, by definition of $\mathcal A$, which implies that $M^2$ is constant inside $\mathcal A$. This simplifies the dynamics of $V^\pi$:
\begin{align*}
dV^\pi_t&=dA^2_t \text{ for }(\omega, t)\in\mathcal A
\end{align*}that is, $V^\pi$ is strictly increasing for $(\omega, t)\in\mathcal A$.
   \finproof

\section{Some Examples of Converging Prices}\label{sec::Examples}
We keep the notation of Section \ref{sec::MainResult}.
\subsection{The martingale $M^2$ is null}\label{Example1}
In this case, we can derive the following quadratic co-variation rule:

\begin{Lemma}\label{ThmNARule}
We suppose that $X$ and $Y$ satisfy the hypotheses from the previous section with $M^2\equiv 0$.  If (NFLVR-S) holds then the process:
\[
\int_0^t\ind_{\{d \langle X,Y\rangle \leq 0\}}dY_s
\]is a $(\bG,\bP)$-supermartingale, which is to say:
\[
\int_0^t\ind_{\{d \langle X,Y\rangle \leq 0\}}dV^Y_s
\]is a decreasing process.
\end{Lemma}
 \proof The result
follows as an application of the Theorem \ref{MainThm}.
\finproof

As an example, let us suppose $T$ is constant, that $M^X$ is a continuous martingale with deterministic quadratic variation,  $f$ a deterministic function and $F(t)=\int_0^t f(s)d\langle M^X\rangle_s$. Then, the following are converging prices:
\begin{align*}
X_t=& X_0+M^X_t\\
Y_t=& Y_0+ \int_0^t M^X_s f(s)d\langle M^X\rangle_s +
\int_0^th_sdM^X_s.
\end{align*}
with
\[
h_t= 1+F(T)-F(t)
\]
Consider $F(t)=1-e^{rt}$
for some $r>0$ (and implicitly $f(t)<0$), then the process $h_t=1+
e^{rt}-e^{rT}$ is negative in an interval of the form $[0,S]$ with
$S<T$, provided that $T$ is large enough. By Lemma \ref{ThmNARule}, there are arbitrage
opportunities if the martingale $M^X$ has negative excursions in the
interval $[0,S]$.

Let us now consider the case of a "survival  claim":
$\xi=\ind_{\{\tau\ > T\}}$, i.e., that pays one monetary
unit if some event $\tau$ does not occur before some fixed
maturity $T$. Suppose that for all investors $\tau$ is a totally inaccessible stopping time; it admits a constant $(\bF^X,\bQ^X)$ intensity $\lambda^X$, resp. a constant $(\bF^Y,\bQ^Y)$ intensity $\lambda^Y$. In this case, $X$ (resp. $Y$) is increasing on the stochastic interval $[0,\tau\wedge T)$ and has a downward jump at $\tau$ if $\tau\leq T$.   More precisely:
\begin{align*}
X_t&=\bQ^X(\tau>T|\F^X_t)=\ind_{\{\tau>t\}}e^{-\lambda^X(T-t)} \\
Y_t&=\bQ^Y(\tau>T|\F^Y_t)=\ind_{\{\tau>t\}}e^{-\lambda^Y(T-t)}.
\end{align*}
 (NFLVR-S) holds in this model (for instance $\bQ^m$ is supermartingale measure with $m=\arg\max_{i\in\{X,Y\}}\lambda^i$). This is in line with Lemma \ref{ThmNARule}: $[X,Y]_t=\Delta X_\tau\Delta Y_\tau\ind_{\{\tau\leq
t\}}\geq 0$ and hence
$\langle X,Y\rangle \geq 0$.

Now, consider an alternative of the above example, where  in the filtration $\bF^X$, the stopping time $\tau$  is predictable, but it is totally inaccessible in $\bF^Y$ with constant intensity, i.e., $Y$ is increasing on the stochastic interval $[0,\tau\wedge T)$ and has a downward jump at $\tau$ if $\tau\leq T$ as above. In the filtration $\bG$ the stopping time  $\tau$ is predictable (because it is predictable in $\bF^X\subset \bG$), therefore the price process $Y$ appears to be $\bG$-predictable and of finite variation, in particular $\langle X,Y\rangle\equiv 0$.
 Then, there are arbitrage opportunities: in the filtration $\bG$ there is no change of measure to  make it a supermartingale. An obvious  arbitrage strategy consists in buying $Y$ and selling it just before $\tau$.

\subsection{Investors with similar risk attitudes in the two markets}\label{Ex_SimRisk}

Let us assume that $\bP$ is a martingale measure for both prices $X$ and $Y$ in their own filtration (but not in the filtration $\bG$), that is: $X_t=\E[\xi|\F^X_t]$ and $Y_t=\E[\xi|\F^Y_t]$.

We illustrate with an example of a defaultable asset: $\xi=\ind_{\{\tau>T\}}\mathcal E(B)_T$, the maturity $T$ being fixed. We assume  that $B$ is a Brownian motion and $\tau$,  the default time of the issuer is an exponentially distributed random variable wih parameter $\lambda$, which is independent from the Brownian motion $B$.

We assume  that the following information sets are available for each of the two markets and the insider, respectively, for $t\in[0,T]$:
\begin{align*}
\F^X_t& =\sigma(B_T)\vee\sigma(\tau \wedge s,s\leq t)\\
\F^Y_t& =\sigma(B_s, s\leq t )\vee\sigma(\tau)\\
\G_t&= \sigma(B_s, s\leq t )\vee\sigma(B_T)\vee\sigma(\tau).
\end{align*}
We denote
$$N_t=\ind_{\{\tau\leq t\}}-\lambda (t\wedge \tau),$$ which is an
$\bF^X$-martingale. Also,
we notice that the $\bF^Y$ Brownian motion $B$ is a
semimartingale in the larger filtration $\bG$, namely
 $$B_t=-\int_0^{t\wedge T}\frac{B_T-B_u}{T-u}du +\beta_t$$ with $\beta$ being a $(\bG,\bP)$ Brownian motion.

Easy computations show that:
\[
X_t=\ind_{\{\tau>t\}}e^{-\lambda(T-t)}{\mathcal E}_T(B) ={\mathcal E}_T(B) e^{-\lambda
T}-\int_0^t X_{s-} dN_s,
\]
which is an $\bF^{X}$-martingale. However, in the filtration $\bG$ the process $X$ is predictable and of finite variation. As it is not decreasing, we conclude by Theorem \ref{ThmGenRepr} that $X$ does not fulfil (NFLVR-S). On the other hand, $Y$ is given by the following $\bF^Y$-martingale:
\[
Y_t=\ind_{\{\tau>T\}}\mathcal E(B)_t=\ind_{\{\tau>T\}}+\int_0^t Y_u dB_u
\]
while in the larger filtration $\bG$, the following decomposition holds for $Y$:
\[
Y_t=\ind_{\{\tau>T\}}-\int_0^tY_u\frac{B_T-B_u}{T-u}du+ \int_0^{t}Y_ud\beta_u,
\]
 where the integral $K_t:=\int_0^tY_u\frac{B_T-B_u}{T-u}du$ is well defined. From \cite{jy:faux} this condition is equivalent to  $\int_0^T  \frac{ Y_s }{\sqrt{T-s}}ds <\infty$ and, since $E(Y_t)\leq 1$ one has $\E\left ( \int_0^T  \frac{|Y_s|}{\sqrt{T-s}}ds\right)<\infty$.
This type of model is known for not satisfying (NFLVR). Imposing short sales constraints  for the insider does not prevent the free lunches. The candidate density process for the fundamental supermartingale measure is:
\[
D^*_t=\mathcal E_t\left(\int_0^\cdot \frac{(B_T-B_u)^+}{T-u}d\beta_u\right),
\]  the fact that $\frac{(B_T-B_u)^+}{T-u}$ is not square integrable  prevents it from being a valid change of measure. See, e.g., \cite{aj}, section 4.2.1.

\subsection{Different risk attitudes in the two markets}
We work directly in the filtration $\bG$, generated by two independent $\bP$-Brownian motions $B$ and $\beta$. We denote  $W:=\rho B+\sqrt{1-\rho^2} \beta$ for some $\rho\in[-1,1]$. The two asset prices are supposed to be as follows:
\begin{align*}
X_t&=X_0+B_t\\
Y_t&=\E^{\bQ^Y}[X_T|\G_t],
\end{align*}
with
\[
\frac{d\bQ^Y}{d\bP}\Big|_{\G_t}:=\mathcal E\left(-\int_0^\cdot W^{Y}_udB_u\right)_t,
\]
 and with $W^Y$ satisfying $dW^Y_t=\rho W^Y_t dt + dW_t$.  We consider $T=\inf\{t\geq 0, X_t=0\}\wedge \bar T$  with $\bar T$ non random (so that the price processes are positive).

 Under the above assumptions, the processes $W^Y$, $B^{Y}_t=B_t+\int_0^tW^Y_udu$ and $\beta$ are $\bQ^Y$-Brownian motions. It can be easily computed that $Y$ has the following $(\bG,\bP)$ decomposition:  \[
Y_t=X_0+\int_0^t(1-\rho(T-u))W^Y_udu+\int_0^t(1-\rho(T-u))dB_u-\sqrt{1-\rho^2}\int_0^t(T-u)d\beta_u.
\] Here $\beta$ is a $(\bG,\bP)$-Brownian motion, independent from $B$.

We have  for $t\leq T$:
\begin{align*}
M^1_t&=\int_0^t h_udB_u\text{ with }h_t=1-\rho(T-t)\\
M^2_t&=-\sqrt{1-\rho^2}\int_0^t(T-u)d\beta_u\\
A_t& =\int_0^t \left(\frac{W^Y_u}{h_u}\right)^+d\langle M^1\rangle _u.
\end{align*}
 For simplicity we fix $\bar T=2$.
We can conclude using Theorem \ref{MainThm} that (NFLVR-S) holds true:
\begin{enumerate}
\item If $\rho\leq 0$, then $h>0$ and conditions (C1) and (C2) are trivially satisfied with $\tilde a^2\equiv 0$. Condition (C3) also holds true.
\item If $\rho>0$ then, $\{(\omega,t)|h_t\leq 0\}=[0,\max(0,2-1/\rho)]$. For $\rho\in(1/2,1]$, this interval is not empty and $dA^2_t>0$ whenever $W^Y_t<0$, and these negative excursions of $W^Y$ occur $a.s.$ on every bounded interval. Hewever, there are no arbitrage opportunities in this case neither: all conditions are fulfilled to construct the fundamental supermartingale measure $\bP^*$.
\end{enumerate}

\subsection{Filtering models with vanishing noise}Another class of examples fitting in the framework of converging prices are filtering models where the noise in the observation process is vanishing at a fixed time $T$.

Let us consider a very simple filtering model. As in the previous example, $X$ is a $\bP$-Brownian motion starting at $X_0$.  Suppose that for $t<T$, the information flow $\bF^Y$ is generated by a noisy observation of $X$, modeled by
\[
O_t=\int_0^tf(X_s)ds+W_t,
\]
 $W$ being a Brownian motion independent from $X$. Furthermore, at time $T$ the value $\xi=X_T=Y_T$ can be observed fully. More precisely:
 \begin{align*}
 \F^Y_t&=\sigma(O_s,s\leq t),\text{ for } t<T\\
 \F^Y_T&=\sigma(O_s,s\leq T)\vee\sigma(\xi).
\end{align*}
As usual, we denote $\G_t=\sigma(X_s,s\leq t)\vee\F^Y_t$. Now, we set:
\[
Y_t:=\E[\xi|\F^Y_t].
\]
We denote by $N_t=O_t-\int_0^t\widehat{f(X_s)}ds$ the innovation process, where as usual $\widehat{ f(X)}$ is the $\bF^Y$-optional projection of the process $f(X)$. We obtain for some $\bF^Y$-predictable process $\psi$ and for $t<T$:
\[
Y_t=\E[X_t|\F^Y_t]=X_0+\int_0^t\psi_udN_u.
\]
The process $\psi$ is given by: $\psi_t=\widehat{f(X_t)X_t}-\widehat{X_t}\widehat{f(X_t)}$ (see Theorem 3.35 in \cite{BainCrisan}).  In the filtration $\bG$, replacing $N_t=W_t+\int_0^t(f(X_s)-\widehat{f(X_s)})ds$ we obtain the following representation:
\begin{align*}
Y_t&= X_0+\left(\int_0^t\psi_u (f(X_u)-\widehat{f(X_u)})du+\ind_{\{t\geq T\}}(\xi-Y_{T-})\right)+\int_0^t\psi_udW_u\\
&=X_0+V^Y_t+M^2_t.
\end{align*}
Here we have an example with $M^1\equiv 0$. We can use Theorem
\ref{ThmGenRepr} to deduce that the dynamics of $Y$ is not
compatible with (NFLVR-S): we can write
\[
V^Y_t= J^Y_t+\left\{\int_0^t\psi_u (f(X_u)-\widehat{f(X_u)})du\right\}
\] where $J^Y_t=\ind_{\{t\geq T\}}(\xi-Y_{T-})$ is not a decreasing process.

\appendix
\renewcommand{\thesubsection}{\Alph{subsection}}
\section{Some recalls on measures and increasing processes}\label{A}
For the reader's convenience we gather here some elementary results that were used in the paper.

\begin{Theorem}\label{diffmeasures} Let $\mu^1$ and $\mu^2$ be two finite (possibly signed) measures.
\begin{itemize}
\item[(a)] Assume that both  $\mu^1$ and $\mu^2$ are positive measures. Then, $(\mu^1-\mu^2)$ is a positive measure only if $\mu^2$ is absolutely continuous with respect to $\mu^1$.
\item[(b)] Assume that  $\mu^1\bot \mu^2$ and furthermore $(\mu^1+\mu^2)$ is a positive measure. Then, both  $\mu^1$ and $\mu^2$ are positive measures.
\end{itemize}
\end{Theorem}

\proof
\begin{itemize}
\item[(a)]Suppose that $(\mu^1-\mu^2)$ is a measure on the $\sigma$-algebra $\F$. Then, for all $A\in\F$, $(\mu^1-\mu^2)(A)\geq 0.$ In particular, if $A$ is such that $\mu^1(A)=0$ then:  $(\mu^1-\mu^2)(A)=  \mu^1(A)-\mu^2(A)=-\mu^2(A) \geq 0$, which implies that $\mu^2(A)=0$ (since we also have  $\mu^2(A)\geq 0$ for $\mu^2$  being a positive measure). In other words: $\mu^2 \ll\mu^1$.
\item[(b)] For all $A\in\F$, $(\mu^1+\mu^2)(A)\geq 0$; the orthogonality condition implies that  $\mu^1(A)\in\{(\mu^1+\mu^2)(A),0\}$ and $\mu^2(A)\in\{(\mu^1+\mu^2)(A),0\}$ hence both are positive measures.
\end{itemize}
\finproof

An increasing process can be seen as a random measure on $\bR^+$, $dA_t (\omega)$, whose distribution function is $A_\bullet(\omega)$. Similarly, a process of finite variation can be seen as a
signed random measure, since it can be written as the difference of two increasing processes.
\begin{Proposition}[\cite{JacodShi} p.30]
Let $A$, $B$ be finite variation processes (resp. increasing processes) such that $dB\ll dA$. Then, there exists an optional (resp. nonnegative) process  $H$ such that $B=\int HdA$ up to an evanescent set. If moreover $A$ and $B$ are predictable, one may choose $H$ to be predictable.
\end{Proposition}

\begin{Proposition}[\cite{DelbScha95}]\label{AnnexA3}
Let $A$, $B$ be c\`adl\`ag, predictable processes of finite variation, with $B$ being increasing. Then, there is a predictable process $\varphi$ and a predictable subset $N$ of $\mathbb R_+\times\Omega$ such that:
\[
A=\int\varphi d B+\int \ind_NdA
\]
and:
\[
\int_{\mathbb R_+}\ind_N(u)dB_u=0.
\]
\end{Proposition}

\subsection*{Acknowledgements}
The research of Monique Jeanblanc is supported by Chair Markets in
Transition  (F\'ed\'eration Bancaire Fran\c caise) and  Labex ANR
11-LABX-0019.

\renewcommand{\refname}{References}

\end{document}